\documentclass[twocolumn,prb,showpacs]{revtex4}
\usepackage{graphicx,psfrag}
\usepackage{slashed,amssymb}

\newcommand{\nn}{\nonumber}

\begin{document}
 
\title{
Emergent Chern-Simons excitations due to electron--phonon interaction
}

\author{Andreas Sinner and Klaus Ziegler}

\affiliation{Institut f\"ur Physik, Universit\"at Augsburg, Universit\"atsstr. 1, D-86135 Augsburg, Germany}

\begin{abstract}

We address the problem of Dirac fermions interacting with transversal optical phonons.
A gap in the spectrum of fermions leads to the emergence of the Chern--Simons 
excitations in the spectrum of phonons. We study the effect of those excitations 
on observable quantities: the phonon dispersion, the phonon spectral density, and the Hall conductivity.

\end{abstract}

\pacs{05.60.Gg, 72.10.Di, 72.20.Dp}

\maketitle

\section{Introduction}  
During the last decade there has been considerable progress in experimental studies of the two-dimensional (2D) Dirac gas in various fields. This was initiated by the investigation of exfoliated and epitaxial graphene~[\onlinecite{Geim2005,Guinea2009,Allen2010}], followed later by experiments with silicene~[\onlinecite{Chen2012}] and with a number of chemical compounds, commonly referred to as topological insulators~[\onlinecite{Hasan2010,Qi2011}]. Other realizations are optical hexagonal lattices filled with ultracold atoms~[\onlinecite{Cirac2010,Korepin2013,Lewenstein2013,Esslinger2014,Bloch2013}] and photonic crystals~[\onlinecite{Szameit2010,Szameit2011}]. Besides the linear Dirac spectrum with nodes, particle--particle interactions can play an important role, depending on the specific system.
For instance, the coupling to lattice vibrations (phonons in condensed matter realization~[\onlinecite{Geim2005,Guinea2009,Allen2010,Chen2012,Hasan2010,Qi2011,Sasaki2008}], shaking in optical lattices~[\onlinecite{Cirac2010,Korepin2013,Lewenstein2013,Esslinger2014,Bloch2013}] 
and photonic crystals with Dirac spectrum~[\onlinecite{Szameit2010,Szameit2011}]) can lead to instabilities~[\onlinecite{Fuchs2007,Ziegler2011a}]. 
Such an instability was recently observed in graphene~[\onlinecite{Politano2015}]. The complexity of the phonon modes can give rise to the plethora of other interesting effects. This is related to the fact that in--plane optical phonons act like effective gauge fields that couple to the Dirac particles. Moreover, it has been known for a long time that the coupling of massive Dirac particles to gauge fields leads to Chern--Simons
excitations~[\onlinecite{Redlich1984,Fradkin1994,Kondo1995,Kimura1994,Barci2000,Dunne1999,Pachos2013}].
It is crucial to note that the Chern--Simons term is created by the expansion of the fermion determinant with respect to a vector field,
where the latter is usually a gauge field. We will show in this paper that an optical in--plane phonon mode, which is a massive vector field, 
can also create such a term. In condensed matter systems there is no ambiguity from the UV regularization because of
a natural momentum cut-off from the underlying lattice.
The main objective of this paper is to study how the Chern--Simons term affects the transport properties. In particular, we will focus 
on the quantum Hall effect and study the Hall conductivity when the Dirac fermions are gapped.

\section{the model} 

\begin{figure*}[t]
\includegraphics[width=9.0cm]{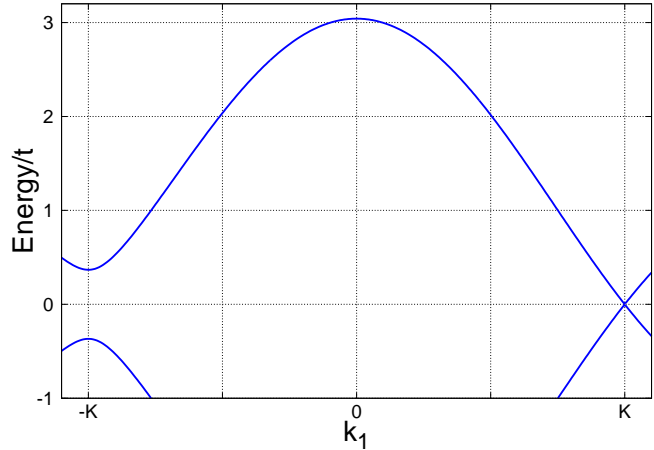}
\caption{(Color online) The fermion band structure of the considered model. The parity symmetry between both fermion copies is explicitly broken as visualized by the gap $2m\sim t$ at the Fermi point in the negative momentum -K versus the gapless state at the Fermi point at the positive momentum  K$={4\pi}/{3\sqrt{3}}$. The total band width $2\Lambda^{}_0\sim6.5t$ is only slightly larger than that of the pristine tight--binding Hamiltonian in Eq.~(\ref{eq:TBH}).}
\label{fig:Model}
\end{figure*}

Our microscopic lattice model describes the interaction between spinless fermions  and monochromatic transversal optical phonons on a two--dimensional hexagonal lattice. The dynamics of the free  fermionic quasiparticles is given by the tight--binding Hamiltonian
\begin{equation}
\label{eq:TBH} 
H^{}_0 = -t\sum_{\langle rr^\prime\rangle}~(c^\dag_r d^{}_{r^\prime} + d^\dag_{r^\prime} c^{}_{r} ),
\end{equation}
where $c$ and $d$ denote fermionic species identified with each sublattice of the hexagonal lattice and $t\sim2.8$eV is the hopping amplitude. We use the unit system with lattice constant, elementary electric charge, and $\hbar=1$. The index $\langle rr^\prime\rangle$ suggests the summation over next neighbors, while the absence of the parenthesis means summation over each sublattice index separately. The Hamiltonian is readily diagonalized by the Fourier--transform and yields a well--known spectrum with two degenerated parity symmetric Fermi points. As a striking feature, close to those Fermi points the fermion dispersion is linear and therefore describes massless Dirac particles. In order to model the fermion mass we have to account either for a staggered potential altering its sign from one lattice site to another~[\onlinecite{Ludwig1994}], or for a spin--orbit coupling term~[\onlinecite{Kane2006}]. 
The parity between the cones is broken by exposing the system to the chemical modification, e.g. by a 
periodic flux ~[\onlinecite{Haldane1988}] or a spin texture~[\onlinecite{Hill2011}]. Combining fermionic operators to  spinors $\Psi^\dag_r = (c^\dag,d^\dag)^{\rm}_r$ and $\Psi^{}_r = (c,d)^{\rm T}_r$, the mass term reads
\begin{equation}
\label{eq:MH}
H^{}_1 =  \mu\sum_r~\Psi^\dag_r \sigma^{}_3 \Psi^{}_r + t^\prime\sum_{r.r^\prime} \Psi^\dag_r \chi^{}_{rr^\prime}\Psi^{}_{r^\prime},
\end{equation}
where $t^\prime$ is the hopping amplitude between second nearest neighbors ($t^\prime/t\sim0.1$~[\onlinecite{Guinea2009}]) and the matrix element of the Haldane term is given by~[\onlinecite{Haldane1988,Hill2011}]
\begin{eqnarray}
\chi^{}_{r r^\prime} = 
\sum^{}_{i=1,2,3}
\Big( 
e^{i\phi}\delta^{}_{r^\prime,r+{a}^{}_i} +
e^{-i\phi}\delta^{}_{r^\prime,r-{a}^{}_i}
\Big),
\end{eqnarray}
with the adjustable Peierls phase $\phi$, and $a^{}_i$ denoting the positions of the second nearest neighbors on the hexagonal lattice. A particular choice of both parameters $\mu=3t^\prime(1+\sqrt{3})/\sqrt{2}$ and $\phi=\pi/4$ opens up a gap at one Dirac point 
with size $2m\sim2\times0.75t$, while keeping the other Dirac point gapless. This band structure is visualized in Fig.~\ref{fig:Model}. 
Moreover, changing the value of the Peierls phase, it is also possible to get into the situation where all cones are gapped with different signs of the Dirac mass.  For the chosen parameter set, all elements of the conductivity tensor of the Haldane model are nonzero~[\onlinecite{Haldane1988,Hill2011}], where the gapped (gapless) channel contributes to the Hall (longitudinal) conductivity. But since we are primarily interested in the anomalous Hall conductivity, we focus on the gapped channel only. Recently, an experimental realization of the Haldane model was reported in Ref.~[\onlinecite{Esslinger2014}].

In our model the monochromatic phonons appear as fluctuating bonds between neighboring sites~[\onlinecite{Ziegler2011,Stauber2007,Mousavi2012}]. The corresponding Holstein Hamiltonian reads
\begin{equation}
\label{holstein}
H^{}_2 = \sum_{\langle rr^\prime\rangle} \{\gamma b^\dag_{rr^\prime}b^{}_{r^\prime r}+
\alpha(b^\dag_{rr^\prime}c^\dag_r d^{}_{r^\prime} + b^{}_{r^\prime r}d^\dag_{r^\prime}c^{}_r)\}.
\end{equation}
The operator $b^{}_{rr^\prime}$ ($b^\dag_{r^\prime r}$) destroys (creates) a phonon between sites $r$ and $r^\prime$. In the continuous limit the phonons couple to the spinors via combinations of non--diagonal Pauli matrices $\sigma^{}_1\pm i\sigma^{}_2$ with the coupling strength $\alpha$. While the phonon frequency $\gamma\sim160-170$meV is intrinsic to the hexagonal lattice~[\onlinecite{Politano2015,Maultzsch2004,Grueneis2009}], the electron--phonon coupling strength $\alpha$, measured in units $\rm energy\cdot{length}$, can be varied in the experiment. 

In the continuous limit we use the low--energy approximation for massive Dirac fermions. This approximation is well known from the literature~[\onlinecite{Wallace1947,Semenoff1984}] and we do not discuss it here. Next, we introduce the coherent state functional integral with the partition function given by 
\begin{equation}
{\cal Z}=\int{\cal D}[\bar\psi,\psi,A]~\exp\{-{\cal S}[\bar\psi,\psi,A]\}
\end{equation}
with the 2+1--dimensional Euclidean action
\begin{eqnarray}
\nn
&\displaystyle {\cal S}[\bar\psi,\psi,A] = \frac{1}{2g}\int d^3x~A^2_\mu &\\
\label{eq:Thirring2}
&\displaystyle + \int d^3x ~\bar\psi[\slashed\partial + m + \frac{i}{\sqrt{2}} A^{}_\mu\gamma^{}_\mu]\psi, &
\end{eqnarray}
where $\slashed\partial = \gamma^{}_0\partial^{}_\tau+\gamma^{}_1\partial_{r^{}_1}+\gamma^{}_2\partial^{}_{r^{}_2}$, $\gamma = (\gamma_0,\gamma_1,\gamma_2)=(\sigma_3,-i\sigma_1\sigma_3,-i\sigma_2\sigma_3)$, $\bar\psi = \psi^\dag\gamma^{}_0$, and $\int d^3x=\int d\tau d^2r$, $\tau$ denoting the imaginary time. The Fermi velocity  $v_F=\sqrt{3}t/2$ is removed from the fermionic part by rescaling $r^{}_i\to v^{}_F r^{}_i$ (not $\tau$) and $\psi\to\psi/v^{}_F$. Real valued components of the phonon field $A_\mu$, $\mu=1,2$ are obtained via
\begin{equation}
b^\dag \to -\frac{A^{}_1 - iA^{}_2}{\sqrt{2}\alpha},\;\; b \to -\frac{A^{}_1 + iA^{}_2}{\sqrt{2}\alpha},
\end{equation}
i.e., by changing from the $U(1)$-- to the $O(2)$--representation of the phonon sector. It is important to stress that, in contrast to gauge fields, our $A^{}_\mu$ are massive fluctuations of the lattice, which cannot be constrained by gauge fixing. Formally however, the classical equation of motion of the field $A$ has approximately the form of the Lorenz gauge condition, cf. Appendix~\ref{app:Lorenz}.

The coupling parameter $g$ in Eq.~(\ref{eq:Thirring2}) is related to the electron-phonon coupling $\alpha$ in (\ref{holstein}) through 
$2g=\alpha^2/(v^2_F\gamma)$. Finally, the slow lattice dynamics ($\sim\partial_\tau A_\mu$) is neglected in comparison to the much faster electron dynamics. Integrating out the phonon fields $A^{}_\mu$ yields an interaction term for the fermions $({g}/{4})(\bar\psi\gamma^{}_\mu\psi)^2$.
This resembles the standard Thirring current--current  interaction~[\onlinecite{Redlich1984,Fradkin1994,Dunne1999,Kondo1995,Kimura1994,Barci2000}] 
with one crucial difference, though, that we have only two spatial currents here. Due to the anticommutativity of the Grassmann field, this interaction can be rewritten as the usual repulsive Hubbard interaction $(g/2)(\psi^\dag\psi)^2$. Throughout the subsequent calculations we implement a large but finite UV--cutoff regularization scheme with only spatial components of the frequency--momentum vector being regularized, while the integration over the Matsubara frequencies stretches to infinity, hence only considering the $T=0$ case here. Fig.~\ref{fig:Model} suggests the appropriate energy cutoff to be $\Lambda^{}_0\sim3.2t\sim10^5 $K.

\section{Phonon theory} 

After integrating the fermions in Eq.~(\ref{eq:Thirring2}), we obtain an effective action in terms of the phonon fields $A$
\begin{equation}
\label{eq:Thirr1}
{\cal S}[A] =\int \frac{A^2_\mu}{2g}d^3x - {\rm tr}\log\left[\slashed\partial + m + i\frac{A^{}_\mu\gamma^{}_\mu}{\sqrt{2}}\right],
\end{equation}
where the functional trace $\rm tr$ consists of the three dimensional integration and the trace with respect to the Dirac space. 
Next we expand the $\rm tr\,log$--term of Eq.~(\ref{eq:Thirr1}) up to second order in the fields $A_\mu$~[\onlinecite{Redlich1984,Fradkin1994,Dunne1999}]:
\begin{eqnarray}
 \nn
&\displaystyle 
- {\rm tr}\log\left[\slashed\partial + m + \frac{i}{\sqrt{2}}A^{}_\mu\gamma^{}_\mu\right] \sim &\\
\nn
&\displaystyle- {\rm tr}\log[\slashed\partial +m] 
-\frac{i}{\sqrt{2}} {\rm tr}[\slashed\partial + m]^{-1}A^{}_\mu\gamma^{}_\mu &\\
&\displaystyle -\frac{1}{4} {\rm tr}[\slashed\partial + m]^{-1}A^{}_\mu\gamma^{}_\mu [\slashed\partial + m]^{-1}A^{}_\nu\gamma^{}_\nu.&
\end{eqnarray}
The linear term in this expansion is traceless, signaling that in classical approximation we get $A_\mu=0$. Thus, Gaussian fluctuations of the phonon field are relevant and lead to the effective action 
\begin{eqnarray}
\nn
&\displaystyle {\cal S}[A] \sim \frac{1}{2g}\int d^3x~A^2_\mu &\\
&\displaystyle - \frac{1}{4}{\rm tr}[\slashed\partial+m]^{-1}A^{}_\mu\gamma^{}_\mu[\slashed\partial+m]^{-1}A^{}_\nu\gamma^{}_\nu, &
\end{eqnarray}
where higher order terms can be neglected because they are irrelevant in terms of a scaling analysis~[\onlinecite{ZinnJustin2002}]. Since the Dirac Hamiltonian is unbounded, some terms in the perturbative expansion are plagued by ultraviolet divergences. Therefore, the fermionic determinant should be properly regularized. There exist many regularization schemes [\onlinecite{Redlich1984,Fradkin1994,Kondo1995,Kimura1994,Barci2000}], 
but in our case the spectrum is naturally bounded by the band of width of $2\Lambda^{}_0\sim6.5t$. Therefore, all terms in the gradient 
expansion acquire small corrections; e.g., of linear or quadratic order in $ m/\Lambda^{}_0$ for the gradient expansion. 
Higher order terms are negligible, as the analysis of the conductivity tensor for the full Haldane model in Ref.~[\onlinecite{Hill2011}] has shown.

Performing a gradient expansion up to second order in the momenta, cf.~Appendix~\ref{app:Loop}, we obtain for the effective action three terms as
\begin{equation}
\label{eq:EffAction}
{\cal S}[A] \sim {\cal S}^{}_\Delta[A] + {\cal S}^{}_{CS}[A] + {\cal S}^{}_{M}[A],
\end{equation}
with the mass term
\begin{equation}
\label{eq:PhonMass} 
 {\cal S}^{}_\Delta[A] = \Delta\int d^3x~A^2_\mu
\ ,
\end{equation}
where
\begin{equation}
\label{eq:PhGap}
\Delta \sim \frac{1}{2g}-\frac{\Lambda^{}_0}{16\pi}
\end{equation}
and $\Lambda^{}_0$ denotes the  UV--cutoff. The second term is the standard Chern--Simons term
\begin{equation}
\label{eq:CSTerm} 
{\cal S}^{}_{CS}[A] \sim -\frac{s\epsilon^{}_{\mu\nu}}{16\pi}\int\frac{d^3P}{(2\pi)^3}~p^{}_0 A^{}_{\mu,P}A^{}_{\nu,-P}, 
\end{equation}
where $\epsilon^{}_{\mu\nu}$ denotes the antisymmetric tensor, $s={\rm sgn}(m)$ and $P=(p^{}_0,p^{}_1,p^{}_2)^{\rm T}$, 
arising due to the energy gap in the spectrum of 2+1--dimensional Dirac fermions. We have neglected corrections ${\cal O}( m^2/\Lambda_0^2)$ in the prefactor of this term (cf. Appendix~\ref{app:Loop}), assuming that the gap (fermion mass) is much smaller than the cutoff momentum~[\onlinecite{Redlich1984,Dunne1999}]. The third term is a Maxwell--like term
\begin{equation}
\label{eq:Maxwell} 
{\cal S}^{}_M[A] \sim \int\frac{d^3P}{(2\pi)^3}~A^{}_{\mu,P}
\frac{\displaystyle P^2\delta^{}_{\mu\nu} - p^{}_\mu p^{}_\nu }{48\pi|m|}A^{}_{\nu,-P}
\ ,
\end{equation}
although its structure is somewhat different from the usual Thirring model~[\onlinecite{Redlich1984,Fradkin1994,Dunne1999,Kondo1995}]. 
As a consequence of the missing gauge invariance of the phonon field, it cannot be written as a product of Maxwell field tensors 
$F^{}_{\mu\nu}=\partial_\mu A_\nu-\partial_\nu A_\mu$. The detailed derivation of Eqs.~(\ref{eq:PhonMass})-(\ref{eq:Maxwell}) is given in 
Appendix~\ref{app:Loop}. The presence of the term (\ref{eq:Maxwell}) in the effective phonon action makes this action
different from the otherwise similar Chern--Simons--Proca action~[\onlinecite{Niemi1994}].

Obviously, the phonon energy gap in Eq.~(\ref{eq:PhGap}) can become negative, signaling an instability in the functional integral. 
This instability is due to the transition into another phase~[\onlinecite{Ziegler2011,Politano2015}].  With the values $\Lambda^{}_0\sim3.2t$ and $m\sim0.75t$, which we have used for Fig.~\ref{fig:Model}, this instability is irrelevant at $t=1$ for sufficiently small electron--phonon coupling:
\begin{equation}
\alpha^2 \lesssim \frac{8\pi}{3}\gamma v^2_F
\ . 
\end{equation}
This inequality is valid in the low--energy approximation given by the action~(\ref{eq:Thirr1}).

\begin{figure*}[t]
\includegraphics[width=10.cm]{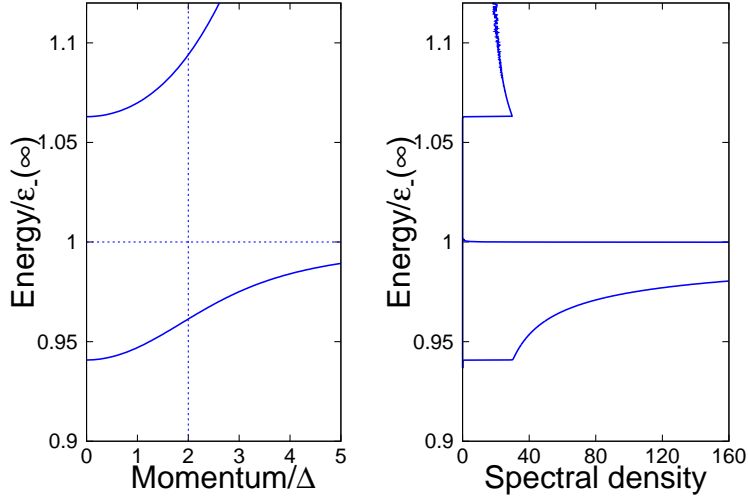} 
\caption{(Color online) Left box: Phonon modes from Eq.~(\ref{eq:Poles}) normalized by $\epsilon^{}_-(\infty)$ depicted versus the momentum in units of the phonon mass, see the discussion in the main text. Right box: The phonon spectral density Eq.~(\ref{eq:SpecFct}) calculated for the same parameter set.}
\label{fig:Spectra}
\end{figure*}

\section{Phonon propagator} 

The Gaussian action in Eqs.~(\ref{eq:EffAction}--\ref{eq:Maxwell}) enables us to calculate 
the correlations of the phonon fields $A^{}_\mu$ without making any additional assumptions. The calculation of the phonon propagator 
is straightforward and yields
\begin{equation}
\label{eq:Corr} 
\langle A^{}_\nu(P)A^{}_\mu(-P)\rangle = \frac{1}{2}\Pi^{}_{\nu\mu}(P),
\end{equation}
where 
\begin{eqnarray}
\nn
&\displaystyle
\Pi(P) = \frac{1}{a^2 [p^2_0+\epsilon^{2}_+(p)][p^2_0+\epsilon^2_-(p)]}
&\\
&\displaystyle
\times\left(
\begin{array}{ccc} 
\label{eq:PhononProp}
\displaystyle \Delta + a(p^2_0 + p^2_1) & & 
\displaystyle sbp^{}_0 + a{p^{}_1p^{}_2} \\
\\
\displaystyle -sbp^{}_0 + ap^{}_1p^{}_2 & &
\displaystyle \Delta + a(p^2_0 + p^2_2) 
\end{array}
\right).
&
\end{eqnarray} 
Here $b=1/16\pi$ denotes the Chern--Simons and $a = 1/(48\pi|m|)$ the Maxwell weight factors. Large (small) values of $|m|$ favor the Chern-Simons (Maxwell-like) term. The elementary excitations of phonons are given by two gapped modes shown in Fig.~\ref{fig:Spectra}, where the dispersion 
relations read 
\begin{equation}
\label{eq:Poles} 
\epsilon^{}_\pm(p) = \sqrt{f(p) \pm t(p)},
\end{equation}
with
\begin{eqnarray}
f(p) &=& \frac{1}{2}\left(p^2 +2\frac{\Delta}{a}+\frac{b^2}{a^2}\right),\\
t(p) &=& \frac{1}{2} \sqrt{\left(p^2+\frac{b^2}{a^2}\right)^2 + 4\frac{\Delta}{a}\frac{b^2}{a^2}},
\end{eqnarray}
and $p^2=p^2_1+p^2_2$. While for $p\ll2\Delta$ both modes grow $\sim p^2$, at $p\gg2\Delta$ the features of non--interacting model are recovered (cf. Fig.~\ref{fig:Spectra}): the lower mode approaches the finite value 
\begin{equation}
\epsilon^{}_-(p\to\infty) = \sqrt{\frac{\Delta}{a}+\frac{b^2}{2a^2}},
\end{equation}
while the upper mode crosses over into the linear regime of Dirac electrons $ \epsilon^{}_+(p\sim\infty)\sim p$. In order 
to understand the role of the Chern--Simons term in the structure of elementary excitations we can formally set $b=0$. In this limit we obtain 
\begin{equation}
\label{eq:MLimit1}
\epsilon^{}_+(p) = \sqrt{p^2+\frac{\Delta}{a}} ,\hspace{4mm} \epsilon^{}_-(p) = \sqrt{\frac{\Delta}{a}},
\end{equation}
which means that the $\epsilon^{}_-$--mode has a flat band and the energy gap between both modes disappears. 
Going even further and setting $\Delta=0$, leaves us with the linear mode of an acoustic phonon.

\section{Hall conductivity} 

A physical quantity, which can be directly evaluated and also experimentally measured, is the Hall conductivity. 
It is calculated from the Kubo formula~[\onlinecite{Ludwig1994}]:
\begin{equation}
\label{eq:Kubo1}
\sigma^{}_{\mu\nu} = \frac{2\pi}{\omega}\int d^3x~e^{-i\omega\tau} \langle(\bar\psi\gamma^{}_\mu\psi)^{}_x(\bar\psi\gamma^{}_\nu\psi)^{}_0\rangle,
\end{equation}
for $\mu\ne\nu$,
where $\langle\cdots\rangle$ denotes the normalized functional integral with the action of Eq.~(\ref{eq:Thirring2}). 
Then in terms of the phonon field the correlation function in  Eq.~(\ref{eq:Kubo1}) becomes:
\begin{equation}
\label{eq:Kubo2} 
\sigma^{}_{\mu\nu} = -\frac{4\pi}{\omega g^2}\int d^3x~e^{-i\omega\tau} \langle  A^{}_{\mu,x}A^{}_{\nu,0} \rangle.
\end{equation}
Some details of this mapping are shown in Appendix~\ref{app:Kubo}. 
With the help of Eq.~(\ref{eq:PhononProp}) we obtain the Hall conductivity as
\begin{equation}
\sigma^{}_{\mu\nu}=\frac{s\epsilon^{}_{\mu\nu}}{a^2g^2}\frac{2\pi b}{[\omega^2+\epsilon^2_+(0)][\omega^2+\epsilon^{2}_-(0)]}.
\end{equation}
Being mainly interested in the behavior of the DC Hall plateaux, we send $\omega\to0$. Since the parameter $s={\rm sgn}(m)$ allows values $\pm1$, there are two plateaux at positive and negative mass value for each combination of $\mu$ and $\nu$. In the non-interacting case the distance between the plateaux is 2$m$. From $\displaystyle \epsilon^2_+(0)\epsilon^{2}_-(0)={\Delta^2}/{a^2}$ we obtain for the system with electron-phonon interaction 
\begin{equation}
\label{eq:dcHall}
\sigma^{dc}_{\mu\nu} = \frac{s\epsilon^{}_{\mu\nu}}{8g^2\Delta^2}
\ ,
\end{equation}
where the definition of $b$ has been used. For weak interaction (or large mass) we approximate $\Delta\sim1/2g$ and reproduce the Hall conductivity of non--interacting massive Dirac fermion gas~[\onlinecite{Ludwig1994,Hill2011,Haldane1988}] 
\begin{equation}
\label{eq:HallCond}
\sigma^{}_H = \frac{s}{2}\frac{e^2}{h}.
\end{equation}
For a finite band width we have $g^2 \Delta^2 \leqslant 1/4$, such that the dc plateaux are shifted away from each other by the electron--phonon
interaction.

\section{Phonon spectral density} 

The spectral density 
\begin{equation}
\label{eq:SpecFct}
S^{}_{\nu\mu}(\omega) = -{\rm Im}\int\frac{d^2p}{(2\pi)^3}~\Pi^{}_{\nu\mu}(p^{}_0\to i\omega + \eta)
\end{equation}
is experimentally accessible in neutron or Bragg scattering experiments.
Below we discuss the diagonal component $S^{}_{\nu\nu}$. The excitation modes shown in Fig.~\ref{fig:Spectra} give a plausible hint to the expected shape of the spectral function: It must comprise two separate structures, corresponding to the two bands and the separation by the gap of size 
$\epsilon^{}_+(0)-\epsilon^{}_-(\infty)$. Moreover, it must reveal an additional energy gap up to $\epsilon_-(0)$. 
By retaining only terms which are resonating at positive frequencies, we obtain 
\begin{eqnarray}
\nn 
&\displaystyle S^{}_{\nu\nu}(\omega)= \frac{1}{a^2}{\rm Im} \int\frac{d^2p}{(2\pi)^3}\times &\\
\label{eq:SpecFct2}
&\displaystyle \left\{\frac{A}{\omega-\epsilon^{}_+-i\eta} + \frac{B}{\omega-\epsilon^{}_--i\eta} \right\},  &
\end{eqnarray}
where the coefficients of the decomposition read 
\begin{eqnarray}
A &=& - \frac{\Delta -a(\epsilon^2_+ -0.5{p^2})}{2\epsilon^{}_+(\epsilon^2_+-\epsilon^2_-)}\\
B &=& ~ \frac{\Delta -a(\epsilon^2_- -0.5{p^2})}{2\epsilon^{}_-(\epsilon^2_+-\epsilon^2_-)}
\ .
\end{eqnarray}
Below we evaluate the spectral function for two extreme parametric regimes, corresponding to the very small and very large fermionic mass $m$.

\begin{figure*}[t]
\includegraphics[width=7cm]{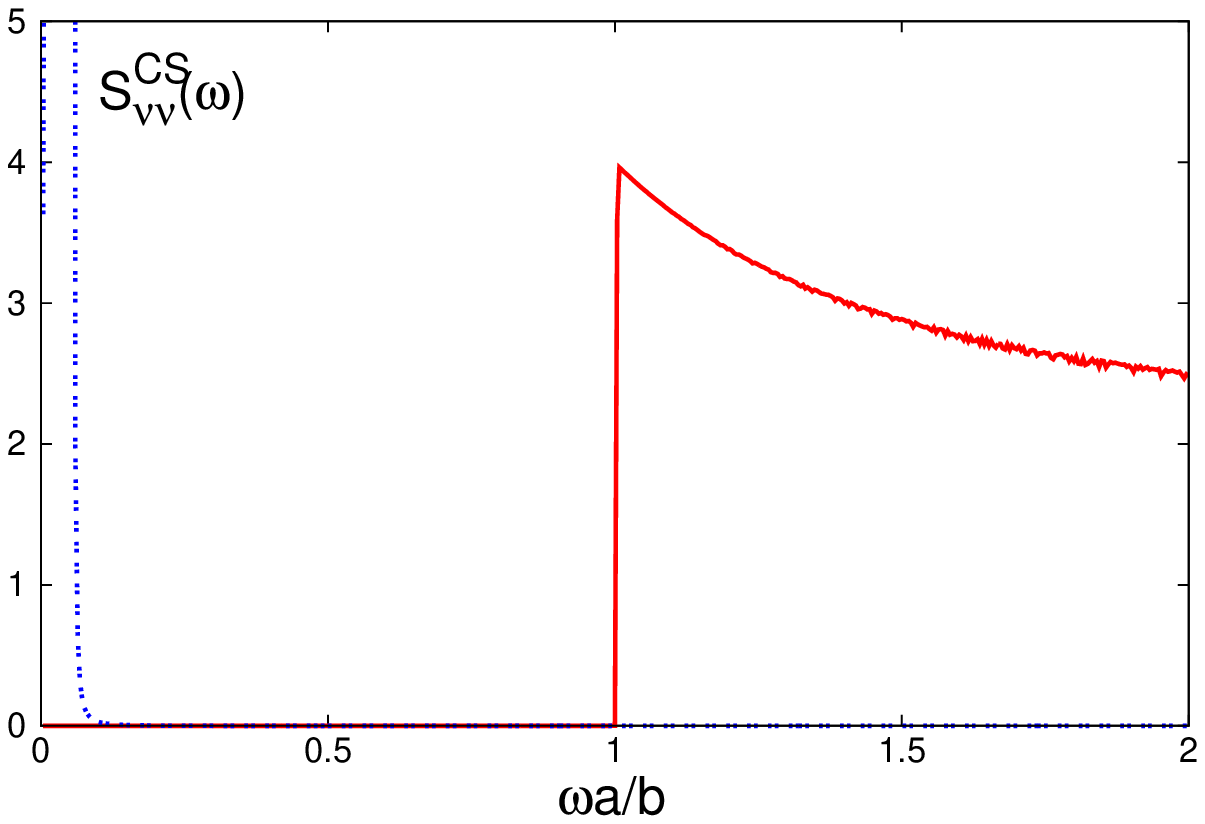}
\hspace{10mm}
\includegraphics[width=7cm]{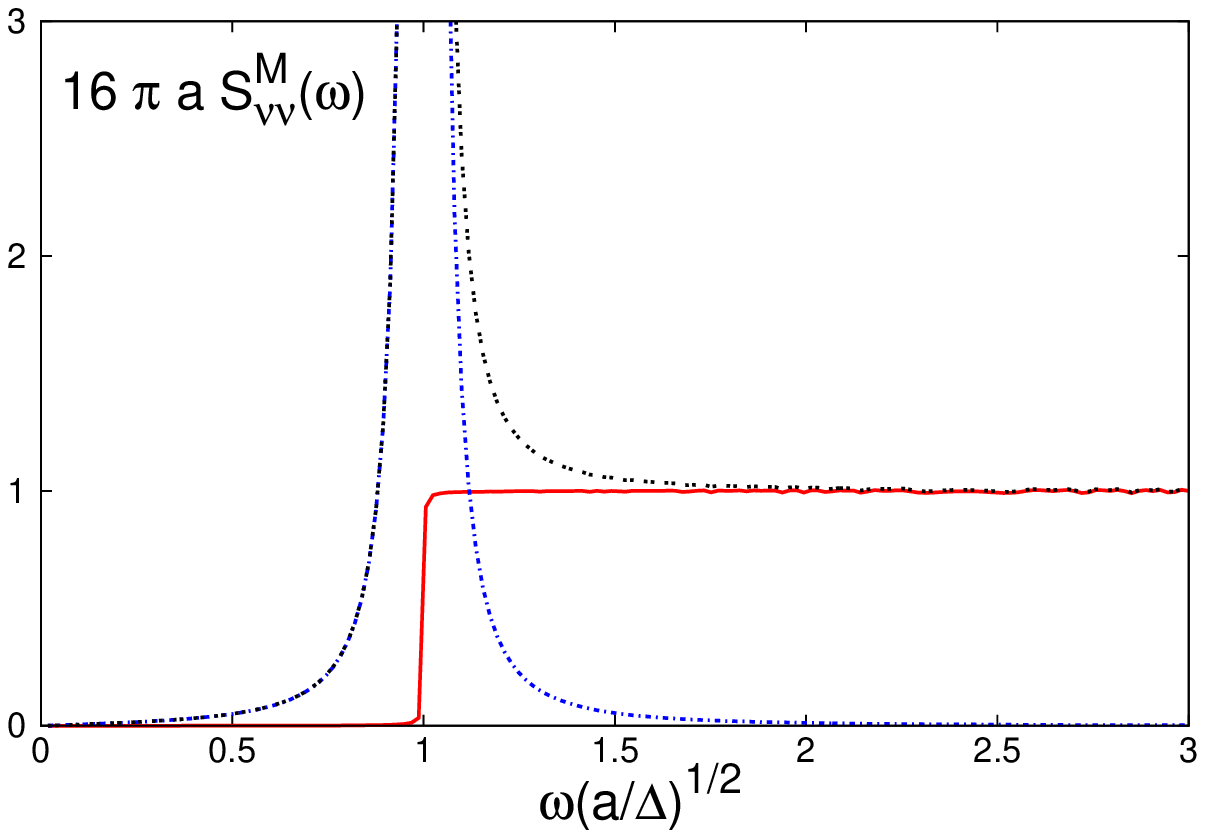}
\caption{(Color online) Left box: The spectral function Eq.~(\ref{eq:SpecFct}) close to the Chern--Simons regime. Blue (dashed) peak on 
the left shows the remnants of the resonance at $\epsilon^{}_-$ which disappears as $a\to0$. It is positioned at $\omega\sim\sqrt{\Delta/a}$.  
Right box: The spectral function close to the Maxwell--like regime. The flat band and continuum contributions are shown in blue (dotted line)  and red (solid line).
The black (dashed) curve is the superposition of the both contributions.}
\label{fig:Limits}
\end{figure*}

{\it Chern--Simons regime:} For large fermionic mass the Chern--Simons contribution is dominant. 
Then for very small $a$ we may approximate $A$ and $B$ as
\begin{equation}
\lim_{a\to0} \frac{A^{}}{a^2} = \frac{1}{2b},\hspace{2mm} \lim_{a\to0}\frac{B^{}}{a^2} = 0 \ .
\end{equation}
After an integration over $p^2$ we obtain
\begin{eqnarray}
\label{eq:CSLimit}
&\displaystyle S^{CS}_{\nu\nu} = \frac{1}{8\pi a}\Theta\left(\omega-\frac{b}{a}\right)\ . &
\end{eqnarray}
Since $b/a=3|m|$, the effect of the Chern--Simons term can be detected as an excitation continuum with frequencies larger than $3|m|$. 
Remarkably, this contribution is indifferent to the value of the phonon gap $\Delta$. For comparison we show in Fig.~\ref{fig:Limits} the 
full numerical evaluation of $S^{}_{\nu\nu}(\omega)$ in the parametric regime close to Eq.~(\ref{eq:CSLimit}).  Actually, the excitation continuum in Eq.~(\ref{eq:CSLimit}) is cut off at a scale related to the band width $\Lambda^{}_0$. This is not included here. 

{\it Maxwell--like regime:} The contribution from the Maxwell--like term is dominant for a small fermionic mass. Then 
\begin{equation}
\lim_{b\to0}\frac{A^{}}{a^2} = \frac{1}{\displaystyle 4a\sqrt{p^2 + \frac{\Delta}{a}}},\hspace{2mm} \lim_{b\to0}\frac{B^{}}{a^2} = 
\frac{1}{4\sqrt{a\Delta}},
\end{equation}
and the energy modes reduce to the expressions in Eq.~(\ref{eq:MLimit1}). Due to the flat band $\epsilon_-$, the integration requires some care. 
In particular, one needs to keep $\eta$ nonzero, which ultimately leads to
\begin{eqnarray}
\nn
S^M_{\nu\nu}(\omega) &=& \frac{1}{8\sqrt{a\Delta}}\frac{\Lambda^2_0}{(2\pi)^2} \frac{\eta}{(\omega-\sqrt{\Delta/a})^2+\eta^2} \\
\label{eq:MLimit} 
&-&\frac{1}{16\pi a} \Theta\left(\sqrt{\Delta/a}-\omega\right),
\end{eqnarray}
where the first sharply peaked contribution is a consequence of the flat band, cf. Fig.~\ref{fig:Limits}.  
Again, the $\Theta$--function which cuts off the excitation continuum at the scale $\sim\Lambda^{}_0$ is not shown in Eq.~(\ref{eq:MLimit}).

\section{Discussion and conclusions} 

Our analysis of massive Dirac particles, which are coupled to optical in-plane phonons,
has revealed an effective phonon model with a Chern--Simons term in Eq. (\ref{eq:CSTerm}) 
and a Maxwell--like term in Eq. (\ref{eq:Maxwell}). These two terms compete, where the former is favored for a large Dirac mass.
The optical phonon dispersion, which is flat without coupling to the Dirac particles, has two branches and becomes parabolic for small momenta
(cf. Fig. \ref{fig:Spectra}). For large momenta one branch is linear and the other is flat. The spectral density of the two modes is shown for large Dirac mass (Chern--Simons regime) in Fig. \ref{fig:Limits} and for small Dirac mass (Maxwell--like regime) in Fig. \ref{fig:Limits}. These results
indicate a separation of the two modes which is obviously caused by the Chern--Simons term. This phenomenon should be observable in inelastic x--ray or neutron scattering experiments in solid state realizations or by laser assisted tunneling experiments on ultracold fermions in optical lattices.  In terms of transport, the Hall conductivity clearly indicates the well--known plateaux $\pm e^2/2h$ for small electron--phonon interaction. For an increasing interaction, though, the plateaux are shifted away from each other. This is a sign that the phonons increase the Hall conductivity. 

A natural question is to ask whether or not the physics we discussed above is related to the fractional quantum Hall effect (FQHE), 
where the Chern--Simons field theory gives a simple and elegant explanation of the phenomenon~[\onlinecite{Dunne1999,Zhang1992,Nagaosa2010}]. 
Within the Ginzburg--Landau approach to FQHE, the Chern--Simons term 
\begin{equation}
\label{eq:CS-FQHE}
{\cal S}^{}_{CS}[a] = \frac{\pi\nu}{2\phi^{}_0} \int d^3x~\epsilon^{}_{\alpha\beta\lambda}a^{}_\alpha  i\partial^{}_{\beta}a^{}_\lambda ,
\end{equation}
of the internal statistical electromagnetic field $a$ with the flux quantum $\phi^{}_0=2\pi$ in chosen units and filling factor $\nu$ appears in the 
action in order to impose the constrain which links $a$ to the particle density. The bosonic Ginzburg--Landau action should be equivalent to a fermionic action, since it is assumed that the quasiparticles are fermions. This requires a statistical field with an odd denominator for the corresponding filling factor:
\begin{equation}
\nu = \frac{1}{2k+1}, 
\end{equation}
i.e. with integer $k$. Including Eq.~(\ref{eq:CS-FQHE}) into the FQHE effective action ultimately leads to the quantized Hall conductivity~[\onlinecite{Zhang1992,Nagaosa2010}]
\begin{equation}
\label{eq:HallCondFQHE}
\sigma^{}_H = \nu\frac{e^2}{h},
\end{equation}
where the SI units are restored. In Eq.~(\ref{eq:CS-FQHE}), the Chern-Simons term appears in a general gauge. It is easy to see that the particular 
gauge choice $a^{}_0=0$ brings it to the form of Eq.~(\ref{eq:CSTerm}), with a different prefactor, though. Clearly, Eqs.~(\ref{eq:HallCond})
and (\ref{eq:HallCondFQHE}) are analogous upon interchanging $\nu$ and $s/2$. This fact suggests that, at least in the large 
mass limit, the phonon field in our model plays a role similar to the statistical vector potential $a$ of FQHE. However, the finite band 
gap of the effective action leads to the corrections of the Hall conductivity given in Eq.~(\ref{eq:dcHall}). 

\vspace{1cm}
\appendix

\section{Classical equation of motion of the phonon field A}
\label{app:Lorenz}

Varying action~(\ref{eq:Thirring2}) with respect to the phonon fields $A$ we acquire
\begin{equation}
\label{eq:App1_1}
A^{}_\mu = - \frac{ig}{\sqrt{2}}~\bar\psi\gamma^{}_\mu\psi = -\frac{ig}{\sqrt{2}}J^{}_\mu,
\end{equation}
where the conserved fermion current density $J^{}_\mu = \bar\psi\gamma^{}_\mu\psi$ obeys the imaginary time continuity condition 
\begin{equation}
\label{eq:App1_2}
\partial^{}_\mu J^{}_\mu = - i\partial^{}_\tau n,
\end{equation}
with $n=\bar\psi\gamma^{}_0\psi$ representing the local electron density. Combination of Eq.~(\ref{eq:App1_1}) and Eq.~(\ref{eq:App1_2}) gives the classical equation of motion (constrain condition) for the phonon field $A$ coupled to the Dirac fermions 
\begin{equation}
\label{eq:App1_3}
\partial^{}_\mu A^{}_\mu + \partial^{}_\tau\varphi \sim 0,
\end{equation}
with $\varphi = gn/\sqrt{2}$, which approximately has the shape of the Lorenz gauge condition. In this paper we concentrate on quantum fluctuations around this classical saddle point for which the condition (\ref{eq:App1_3}) is not valid.

\begin{widetext}
\section{Evaluation of the phonon loop}
\label{app:Loop}

Below we perform the gradient expansion in the action Eq.~(\ref{eq:EffAction}) which leads to Eqs.~(\ref{eq:PhonMass}), (\ref{eq:CSTerm}) and (\ref{eq:Maxwell}). 
The second term from Eq.~(\ref{eq:EffAction}) reads:
\begin{eqnarray}
&\displaystyle \frac{1}{4}{\rm tr}[\slashed\partial+m]^{-1}A^{}_\mu\gamma^{}_\mu[\slashed\partial+m]^{-1}A^{}_\nu\gamma^{}_\nu = 
\int\frac{d^3P}{(2\pi)^3}~A^{}_{\mu,P}A^{}_{\nu,-P}
~ \frac{1}{4}{\rm Tr}\int\frac{d^3Q}{(2\pi)^3}~G(Q)\gamma^{}_\mu G(Q+P)\gamma^{}_\nu, &
\end{eqnarray}
with $G(Q)=[-i\slashed Q + m]^{-1}$. We evaluate the loop function by the method of Feynman parameter, using
\begin{equation}
\frac{1}{AB} = \int^1_0 dx~\frac{1}{[(1-x) A + xB]^2}.
\end{equation}
This gives
\begin{eqnarray}
\nn
\frac{1}{4}{\rm Tr}\int\frac{d^3Q}{(2\pi)^3}~
\frac{[i\slashed Q + m]\gamma^{}_\mu[i(\slashed Q+\slashed P)+m]\gamma^{}_\nu}
{\displaystyle [Q^2 + m^2][(Q+P)^2+m^2]}  = 
\frac{1}{4}{\rm Tr}\int_0^1 dx\int\frac{d^3Q}{(2\pi)^3}~
\frac{[i\slashed Q + m]\gamma^{}_\mu[i(\slashed Q+\slashed P)+m]\gamma^{}_\nu}
{[(1-x)Q^2 + x(Q+P)^2 + m^2]^2}.
\end{eqnarray}
Shifting $Q\to Q - x P$ symmetrizes the denominator and all terms containing odd powers of $Q$ in the numerator vanish due to angular integration. The remaining expression is
\begin{equation}
\frac{1}{4}{\rm Tr}\int_0^1 dx~\int\frac{d^3Q}{(2\pi)^3}~
\frac{-im\slashed P\gamma^{}_\mu\gamma^{}_\nu + m^2\gamma^{}_\mu\gamma^{}_\nu - \slashed Q\gamma^{}_\mu\slashed Q\gamma^{}_\nu + x(1-x)\slashed P\gamma^{}_\mu\slashed P\gamma^{}_\nu}
{[Q^2+m^2+x(1-x)P^2]^2},
\end{equation}
Next we perform the trace. The Chern-Simons term appears from 
$$
-im{\rm Tr}\{\slashed P\gamma^{}_\mu\gamma^{}_\nu\} = -imp^{}_\alpha{\rm Tr}(\gamma^{}_\alpha\gamma^{}_\mu\gamma^{}_\nu) = 2mp^{}_0\epsilon^{}_{\mu\nu},
$$
since $\mu,\nu=1,2$. Next we have
$$
-{\rm Tr}(\slashed Q\gamma^{}_\mu\slashed Q\gamma^{}_\nu) = - q^2_0 {\rm Tr}(\gamma^{}_0\gamma^{}_\mu\gamma^{}_0\gamma^{}_\nu)-q^{}_\alpha q^{}_\beta{\rm Tr}(\gamma^{}_\alpha\gamma^{}_\mu\gamma^{}_\beta\gamma^{}_\nu) 
\to 2\delta^{}_{\mu\nu}q^2_0 - \frac{1}{2}q^2{\rm Tr}(\gamma^{}_\alpha\gamma^{}_\mu\gamma^{}_\alpha\gamma^{}_\nu)  = 2\delta^{}_{\mu\nu}q^2_0,
$$ 
where we made use of the angular averaging 
$
\displaystyle\int_0^{2\pi}\frac{d\phi}{2\pi}~q^{}_\alpha q^{}_\beta = \frac{1}{2} q^2 \delta^{}_{\alpha\beta}\int_0^{2\pi}\frac{d\phi}{2\pi},
$
and of the fact that 
$\displaystyle
\sum_{\alpha=1,2} \gamma^{}_{\alpha}\gamma^{}_\mu\gamma^{}_\alpha\gamma^{}_\nu = \left.\gamma^{}_\mu\gamma^{}_\nu\right|_{\alpha=\mu}-\left.\gamma^{}_\mu\gamma^{}_\nu\right|_{\alpha\neq\mu} = 0.
$
The last term simplifies to 
\begin{eqnarray}
\nn
&\displaystyle
{\rm Tr}\slashed P\gamma^{}_\mu\slashed P\gamma^{}_\nu = -2p_0^2\delta^{}_{\mu\nu} + p^{}_\alpha p^{}_\beta {\rm Tr}(\gamma^{}_\alpha\gamma^{}_\mu\gamma^{}_\beta\gamma^{}_\nu) = -2 p^2_0\delta^{}_{\mu\nu} + p^{}_\alpha p^{}_\beta{\rm Tr} (\delta^{}_{\alpha\mu} + i\epsilon^{}_{\alpha\mu\tau}\gamma^{}_\tau)(\delta^{}_{\beta\nu}+i\epsilon^{}_{\beta\nu\eta}\gamma^{}_\eta)&\\
&\displaystyle
=  -2 p^2_0\delta^{}_{\mu\nu} + 2p^{}_\alpha p^{}_\beta (-\delta^{}_{\alpha\beta}\delta^{}_{\mu\nu} + \delta^{}_{\alpha\mu}\delta^{}_{\beta\nu} + 
\delta^{}_{\alpha\nu}\delta^{}_{\beta\mu})  = -2P^2\delta^{}_{\mu\nu} + 4p^{}_\mu p^{}_\nu.
&
\end{eqnarray}
Hence, prior to the momentum integration we have
\begin{equation}
\frac{1}{2}\int_0^1 dx~\int\frac{d^3Q}{(2\pi)^3}~
\frac{(m^2+q^2_0)\delta^{}_{\mu\nu} +  mp^{}_0\epsilon^{}_{\mu\nu} - x(1-x)(P^2\delta^{}_{\mu\nu}-2p^{}_\mu p^{}_\nu)}
{[Q^2+m^2+x(1-x)P^2]^2},
\end{equation}
and the gradient expansion can be easily performed. To the second order in momenta it reads
\begin{eqnarray}
\nn
&\displaystyle
\frac{1}{2}\int\frac{d^3Q}{(2\pi)^2}~\left\{ \frac{\delta^{}_{\mu\nu}(m^2+q^2_0)}{(Q^2+m^2)^2} + 
\frac{\epsilon^{}_{\mu\nu}mp^{}_0}{(Q^2+m^2)^2} -
\int_0^1dx~x(1-x) ~ \left(
\frac{2\delta^{}_{\mu\nu}(m^2+q^2_0)P^2}{[Q^2+m^2]^3} + \frac{(P^2\delta^{}_{\mu\nu}-2p^{}_\mu p^{}_\nu)}{[Q^2+m^2]^2}
\right)
\right\}
&
\\
&\displaystyle
\sim \frac{\delta^{}_{\mu\nu}}{16\pi}\frac{\Lambda^{2}_0}{\sqrt{\Lambda^2_0 + m^2}} 
+ {\rm sgn}(m) {\epsilon^{}_{\mu\nu}}{\cal C} p^{}_0
- (P^2\delta^{}_{\mu\nu} - p^{}_\mu p^{}_\nu)\left[\frac{1}{48\pi |m|} - {\cal O}\left(\frac{m^2}{\Lambda^2_0}\right)\right], 
&
\end{eqnarray}
where
$\displaystyle
{\cal C} = 1 - \left[1 + \left(\frac{\Lambda^{}_0}{m}\right)^2\right]^{-1/2},
$
which gives the loop contributions to the phonon mass Eq.~(\ref{eq:PhonMass}) and both momentum dependent terms Eq.~(\ref{eq:CSTerm}) and (\ref{eq:Maxwell}). For $\Lambda^2_0\gg m^2$ we get to Eqs.~(\ref{eq:EffAction}-\ref{eq:Maxwell}).

\section{Kubo formula in phonon representation}
\label{app:Kubo}

Effective actions Eq.~(\ref{eq:Thirring2}) and Eq.~(\ref{eq:Thirr1}) are equivalent to 
\begin{eqnarray}
\label{eq:ActN1}
{\cal S}[\bar\psi,\psi,A] = \frac{1}{2g}\int d^3x~\left[A^{}_\mu+\frac{ig}{\sqrt{2}}(\bar\psi\gamma^{}_\mu\psi)\right]^2 
+ \int d^3x\left\{ \bar\psi[\slashed\partial+m]\psi + \frac{g}{4}(\bar\psi\gamma^{}_\mu\psi)^2 \right\}.
\end{eqnarray}
The fermion average in the Kubo formula Eq.~(\ref{eq:Kubo1}) can be rewritten as
\begin{eqnarray}
\nn
\langle(\bar\psi^{} \gamma^{}_{\mu}\psi^{})^{}_x(\bar\psi \gamma^{}_{\nu}\psi)^{}_0\rangle &=& 
-\frac{2}{g^2} \langle\left(-A^{}_{\mu}+A^{}_{\mu} + \frac{ig}{\sqrt{2}}\bar\psi^{} \gamma^{}_{\mu}\psi^{}_n\right)^{}_x
\left(-A^{}_{\nu} + A^{}_{\nu} + \frac{ig}{\sqrt{2}}\bar\psi^{} \gamma^{}_{\nu}\psi^{} \right)_0 \rangle \\
\nn
&=& -\frac{2}{g^2}\langle 
A^{}_{\mu,x}A^{}_{\nu,0} +
\left(A^{}_{\mu} + \frac{ig}{\sqrt{2}}\bar\psi^{}\gamma^{}_{\mu}\psi^{}\right)_x 
\left(A^{}_{\nu} + \frac{ig}{\sqrt{2}}\bar\psi^{}\gamma^{}_{\nu}\psi^{}\right)_0
\rangle \\
\nn
&& +\frac{2}{g^2}\langle 
 A^{}_{\mu,x}\left(A^{}_{\nu} + \frac{ig}{\sqrt{2}}\bar\psi^{} \gamma^{}_{\nu}\psi^{} \right)_0 + 
A^{}_{\nu,0}\left(A^{}_{\mu} + \frac{ig}{\sqrt{2}}\bar\psi^{} \gamma^{}_{\mu}\psi^{} \right)_x
\rangle. 
\end{eqnarray}
Keeping track on Eq.~(\ref{eq:ActN1}) and assuming $A^{}_\mu+\frac{ig}{\sqrt{2}}(\bar\psi\gamma^{}_\mu\psi)$ to be an independent integration variable  we realize, that the mixed terms with both phonons and fermions vanish for $x\neq 0$, since action (\ref{eq:ActN1}) is diagonal in space. Then, the functional fermion integration in~(\ref{eq:ActN1}) can be performed which yields for the only remaining part Eq.~(\ref{eq:Kubo2}). 
\end{widetext}


\begin{thebibliography}{99}
%
\bibitem{Geim2005} K. S. Novoselov, A. K. Geim, S. V. Morozov, D. Jiang, M. I. Katsnelson, I. V. Grigorieva, S. V. Dubonos, and A. A. Firsov, Nature {\bf 438}, 197 (2005).
%
\bibitem{Guinea2009} A. H. Castro Neto, F. Guinea, N. M. R. Peres, K. S. Novoselov, and A. K. Geim, Rev. Mod. Phys. {\bf 81}, 109 (2009).
%
\bibitem{Allen2010} M. J. Allen, V. C. Tung, and R. B. Kaner, Chem. Rev. {\bf 110}, 132 (2010).
%
\bibitem{Chen2012} L. Chen, Ch.-Ch. Liu, B. Feng, X. He, P. Cheng, Z. Ding, Sh. Meng, Y. Yao, and K. Wu, Phys. Rev. Lett. {\bf 109}, 056804 (2012).
%
\bibitem{Hasan2010}  M. Z. Hasan and C. L. Mele, Rev. Mod. Phys. {\bf 82}, 3045 (2010).
%
\bibitem{Qi2011} X.-L. Qi and S.-C. Zhang, Rev. Mod. Phys. {\bf 83}, 1057 (2011).
%
\bibitem{Sasaki2008} K. Sasaki and R. Saito, Prog. Theor. Phys. Suppl., {\bf 176}, 253 (2008).
%
\bibitem{Cirac2010} J. I. Cirac, P. Maraner, and J. K. Pachos, Phys. Rev. Lett. {\bf 105}, 190403 (2010).
%
\bibitem{Korepin2013} D. G. Angelakis, M.-X. Huo, D. Chang, L. C. Kwek, and V. Korepin, Phys. Rev. Lett. {\bf 110}, 100502 (2013).
%
\bibitem{Lewenstein2013} J. Struck, M. Weinberg, C. \"Olschläger, P. Windpassinger, J. Simonet, K. Sengstock, R. H\"oppner, P. Hauke, A. Eckardt, M. Lewenstein, and L. Mathey, Nat. Phys. {\bf 9}, 738 (2013).
%
\bibitem{Bloch2013} M. Aidelsburger, M. Atala, M. Lohse, J. T. Barreiro, B. Paredes, and I. Bloch, Phys. Rev. Lett. {\bf 111}, 185301 (2013).
%
\bibitem{Esslinger2014} G. Jotzu, M. Messer, R. Desbuquois, M. Lebrat, T. Uehlinger, D. Greif, and T. Esslinger, Nature (London) {\bf 515}, 237 (2014).
%
\bibitem{Szameit2010} F. Dreisow, M. Heinrich, R. Keil, A. T\"unnermann, S. Nolte, S. Longhi, and A. Szameit, Phys. Rev. Lett. {\bf 105}, 143902 (2010).
%
\bibitem{Szameit2011} A. Szameit, M. C. Rechtsman, O. Bahat-Treidel, and M. Segev, Phys. Rev. A {\bf 84}, 021806(R), (2011).
%
\bibitem{Fuchs2007} J.-N. Fuchs and P. Lederer, Phys. Rev. Lett. {\bf 98}, 016803 (2007).
%
\bibitem{Ziegler2011a} K. Ziegler, E. Kogan, E. Majernikova and S. Shpyrko, Phys. Rev. B {\bf 84}, 073407 (2011).
%
\bibitem{Politano2015} A. Politano, F. de Juan, G. Chiarello, and H. A. Fertig, Phys. Rev. Lett. {\bf 115}, 075504 (2015).
%
\bibitem{Dunne1999} G. Dunne, {\it Aspects of Chern-Simons Theory}, in {\it Topological aspects of low dimensional systems}, 
edited by A. Comtet, T. Jolicoeur, S. Ouvry, and F. David, Les Houches Summer School, (Springer, Berlin, Heidelberg, 1999),  Vol. {\bf 69}, 177--263.
%
\bibitem{Redlich1984} A. N. Redlich, Phys. Rev. Lett. {\bf 52}, 18 (1984); Phys. Rev. D {\bf 29}, 2366 (1984).
%
\bibitem{Fradkin1994} E. Fradkin and F. Schaposnik, Phys. Lett. B {\bf 338}, 253 (1994).
%
\bibitem{Kondo1995} K.-I. Kondo, Prog. Theo. Phys. {\bf 94}, 899 (1995).
%
\bibitem{Kimura1994} T. Kimura, Prog. Theo. Phys. {\bf 92}, 899 (1994).
%
\bibitem{Barci2000} D. G. Barci, J. F. Medeiros Neto, L. E. Oxman, S. P. Sorella, Nucl. Phys. B{\bf 600}, 203 (2001).
%
\bibitem{Pachos2013} G. Palumbo and J. K. Pachos, Phys. Rev. Lett. {\bf 110}, 211603 (2013).
%
\bibitem{Ludwig1994} A. W. W. Ludwig, M. P. A. Fisher, R. Shankar, and G. Grinstein,
Phys. Rev. B {\bf 50}, 7526 (1994).
%
\bibitem{Kane2006} C. L. Kane and E. J. Mele, Phys. Rev. Lett. {\bf 95}, 226801 (2005).
%
\bibitem{Haldane1988} F. D. M. Haldane, Phys. Rev. Lett. {\bf 61}, 2015 (1988).
%
\bibitem{Hill2011} A. Hill, A. Sinner and K. Ziegler, N. J. Ph. {\bf 13}, 035023 (2011).
%
%
\bibitem{Ziegler2011} K. Ziegler and E. Kogan, Eur. Phys. Lett. {\bf 95}, 36003 (2011).
%
\bibitem{Stauber2007}  T. Stauber and N. M. R. Peres, J. Phys.: Condens. Matter {\bf 20}, 055002 (2008).
%
\bibitem{Mousavi2012} H. Mousavi, Comm. Theor. Phys. {\bf 57}, 482 (2012).
%
\bibitem{Maultzsch2004} J. Maultzsch, S. Reich, C. Thomsen, H. Requardt, and P. Ordej\'{o}n, Phys. Rev. Lett. {\bf 92}, 075501 (2004).
%
\bibitem{Grueneis2009} A. Gr\"uneis, J. Serrano, A. Bosak, M. Lazzeri, S. L. Molodtsov, L. Wirtz, C. Attaccalite, M. Krisch, A. Rubio, F. Mauri, and T. Pichler, Phys. Rev. B {\bf 80}, 085423 (2009).
%
\bibitem{Wallace1947} P. K. Wallace, Phys. Rev. {\bf 71}, 622 (1947).
%
\bibitem{Semenoff1984} G. W. Semenoff, Phys. Rev. Lett. {\bf 53}, 2449 (1984).
%
\bibitem{ZinnJustin2002} J. Zinn-Justin, {\it Quantum field theory and critical phenomena}, Claredon, Oxford, UK, (2002). 
%
\bibitem{Niemi1994} A.J. Niemi and V. V. Sreedhar, Phys. Lett. B {\bf 336}, 381 (1994).
%
\bibitem{Zhang1992} S.-C. Zhang, Int. J. Mod. Phys., {\bf 6}, 25 (1992).
%
\bibitem{Nagaosa2010}  N. Nagaosa, {\it Quantum field theory in condensed matter physics}, Springer, Berlin, (2010).
%
%
%
%













\end{thebibliography}
\end{document}